# Two Dimensional Connectivity for Vehicular Ad-Hoc Networks

Masoud Farivar
Department of Electrical Engineering
Sharif University of Technology
Email: farivar@ee.sharif.edu

Behzad Mehrdad
Department of Electrical Engineering
Sharif University of Technology
Email: behzad.mehrdad86@gmail.com

Farid Ashtiani
Advanced Communications Research Institute
Department of Electrical Engineering
Sharif University of Technology
Email: ashtianimt@ee.sharif.edu

*Abstract*—In this paper, we focus on two-dimensional connectivity in sparse vehicular ad hoc networks (VANETs). In this respect, we find thresholds for the arrival rates of vehicles at entrances of a block of streets such that the connectivity is guaranteed for any desired probability. To this end, we exploit a mobility model recently proposed for sparse VANETs, based on BCMP open queueing networks and solve the related traffic equations to find the traffic characteristics of each street and use the results to compute the exact probability of connectivity along these streets. Then, we use the results from percolation theory and the proposed fast algorithms for evaluation of bond percolation problem in a random graph corresponding to the block of the streets. We then find sufficiently accurate two-dimensional connectivity-related parameters, such as the average number of intersections connected to each other and the size of the largest set of inter-connected intersections. We have also proposed lower bounds for the case of heterogeneous network with two transmission ranges. In the last part of the paper, we apply our method to several numerical examples and confirm our results by simulations.

## I. INTRODUCTION

Vehicular ad hoc networks (VANETs) play a key role in future for safe transportation and easy communications among the riders of vehicles. Compared to other types of mobile ad hoc networks (MANETs), VANETs have some distinct features, i.e., roadmap-restricted mobility patterns as well as higher-speed mobility of communications nodes.

Several papers have considered VANETs at different aspects [1]. Among them, connectivity is of crucial importance. The main reason is due to strong dependency of network efficiency and routing algorithms on this aspect. In fact, in a VANET, we have the option of carry and forward approach [2], as well as radio propagation for packet transfer. Since radio propagation is obviously very faster than carry and forward approach on one hand, and the transit delay is of crucial importance in achieving the basic goal of VANETs, i.e., safe transportation, on the other hand, connectivity plays the main role in routing decisions. Succinctly, by knowing the connectivity status at different intersections, routing decisions can be made in a manner in order to attain minimum packet transfer delay.

Up to now, many papers have been focused on connectivity of MANETs. The authors in [3] have considered the connectivity of an ad hoc network with a finite number of nodes uniformly distributed in a one-dimensional network. An exact formula for probability of connectivity for the above network (one-dimensional connectivity), can be found in [4]. In [5], the authors considered one-dimensional random waypoint mobility model for the nodes and then obtained network connectivity. Some of the other researches have been concentrated on the effect of transmission range on connectivity in both homogeneous and heterogeneous situations [6]. Moreover, the authors in [7] have addressed the effect of random power selection against fixed power selection in tradeoff between connectivity and interference. In this respect, [8] has also been focused on the relation between connectivity and interference in a general mobile ad hoc network with spatial Poisson process, based on percolation theory.

Actually, in this paper, two nodes are connected provided that they are in the transmission range of each other and signal to interference ratio exceeds a threshold. So, the results in [8] are very useful when CDMA is the multiple access technique in the network. Recently, a performance evaluation of connectivity for a VANET has been considered in [9] along a highway. And loose lower and upper bounds for connectivity along a street have been reported in [4], recently.

To the best of our knowledge, the papers focusing on VANET connectivity including the above mentioned ones, only consider one-dimensional connectivity, i.e., along a street or highway. But for a typical block of streets including several intersections we need two-dimensional connectivity. In fact, there are many paths between two typical intersections several streets away from each other. When the number of streets between two intersections increases, the connection possibilities increases more and more. Hence, connectivity between two typical intersections becomes more complex.

Another important problem in VANETs is how the behavior of vehicles at different intersections and the arrival rate of vehicles at entrances affect the connectivity. Depending on the resulting arrival rates of vehicles along different streets at both directions, we may have distinct probability of connectivity along different streets. Proposing a computational framework in order to find the relation of these parameters and connectivity has not been reported in the literature. In this paper, we will fill the above gaps. To this end, we exploit a recently proposed mobility model [4] for VANETs. This mobility model is based on a BCMP queueing network, i.e., a quasi-reversible queueing network with nice product-form solution property. By solving the related traffic equations we are able to find the spatial traffic distribution and especially the arrival rate of vehicles at both directions of each street. Then, we find the exact probability of connectivity along each street. Afterwards, we model the streets and intersections as a lattice random graph such that each two consecutive intersections has an edge (bond) with a probability corresponding to the probability of connectivity of that street. Then, with resort to percolation theory and fast computational methods for bond percolation proposed in the literature, we are able to obtain the related results of connectivity among intersections. However, for the case in which we have asymmetric situations among different streets, we obtain lower and upper bounds for connectivity-related parameters. At all cases, we also obtain a threshold for vehicles arrival rate at each entrance in order to have the desired connectivity-related results. We will see that the general behavior and the results predicted by applying percolation theorems on infinite-size lattices are also applicable to finite size VANETs. At the last part of the paper, we consider several numerical results obtained by our approach and confirm our results by simulations.

Following this introduction, we review the mobility model exploited in this paper, in Section II, and find the spatial probability distributions based on vehicle's arrival rate and the behavior of vehicles at intersections. In Section III, we compute the probability of connectivity at a typical street with respect to spatial distribution at both directions. Afterwards we will establish a bound for the case of heterogenous transmission ranges. Then, we make a brief review on percolation results in Section IV. We discuss fast computational method for bond percolation, proposed in the literature in section V. In Section VI, we illustrate the efficiency of our approach in a number of scenarios for both symmetric and asymmetric situations. We also confirm our numerical results by simulation in the same section. We conclude this paper in Section VII.

## II. Review On Mobility Model

As discussed in the previous section, we have exploited a mobility model recently proposed in the literature [4]. Although two mobility models in [4] have been proposed for sparse and dense situations, respectively, we only focus on sparse situations since in dense situations, there is no problem for connectivity. In other words, the dense streets are connected with probabilities almost equal to one, so they do not require any special concern. Thus, in what follows, we discuss the mobility patterns and their modeling, only for sparse scenarios

Obviously, the mobility pattern of each vehicle is strongly affected by the topology of the roads. So, the first step for considering a mobility pattern, reverts to the road topology at the desired region. In this paper, we consider a region similar to the one illustrated in Fig. 1, however, there is not any restriction on the shape and direction of the streets. In the desired topology (Fig. 1), we have several intersections and streets. We assume that the vehicles arrive to the desired region from entrances according to a Poisson process. Since the behaviors of vehicles along the streets are usually different, we differentiate the movement of a vehicle along different parts of a typical street. In this respect, we make a general assumption; a typical vehicle at each street has three phases of movement.

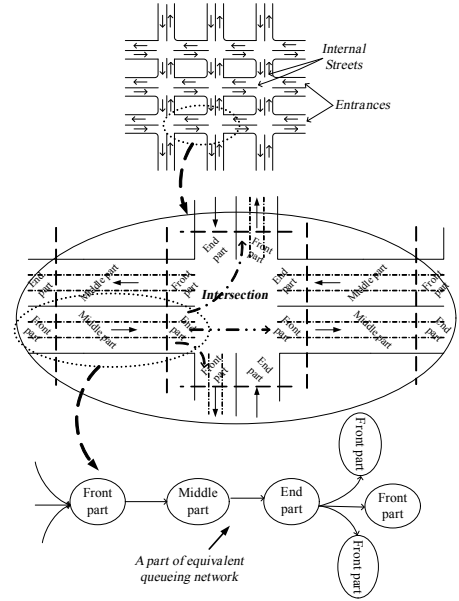

Fig. 1. A typical roadmap and the method of mapping different parts of the streets onto equivalent queueing nodes.

The first phase of movement indicates a transient phase that corresponds to front part of the street. In this phase the vehicle has a changing speed until it reaches to a nearly constant speed. The vehicle maintains its speed at the middle part of the street, equivalent to the second phase of its movement. This constant speed depends on the driving habits of the driver and other factors in the street. By considering the vehicles at that street in a statistical sense, we have assumed three categories of speeds; low, medium, and fast. Each category at the above division has a range of speeds as well as a general distribution (e.g., uniform), such that each vehicle of a typical category selects a speed from the corresponding range according to the corresponding distribution. After moving along the middle part of the street, we have considered another phase of movement due to approaching to an intersection. We have considered two categories of speeds in general for this end part of the streets; low, and medium. Low speeds represent the situation of encountering stoplights, stop-signs, or any other means of blocking. And medium speed category corresponds to the case of not encountering an effective obstacle. Correspondingly, we have considered two speed categories for the vehicles at the front part; one category represents the vehicles arriving at the front part with low speed and the other category represents the vehicles arriving at the front part with medium speed.

Therefore, we have two speed categories at both front and end parts of each street and three speed categories for the vehicles at the middle part of the streets. Usually, we assume that the length of the front and end parts of the streets are very smaller than the length of the middle part. It is worth mentioning that each speed category represents a range of speeds with a specific general distributions (e.g., uniform, truncated normal, etc.).

For the sake of simplicity, in this paper we have assumed that through a street a vehicle does not change its direction except at the intersections. Also, a vehicle at an intersection can go straight forward, turn to the right or turn to the left. However, these assumptions do not limit the considered mobility pattern.

In the above discussed mobility pattern, if the density of the vehicles at a typical street is low, i.e., a sparse scenario, we can consider the mobility of the vehicles independent of each other. Actually, in this scenario when a vehicle approaches to the slower vehicle at its front, the faster one overtakes the slower one with a simple maneuver. Therefore, the independency assumption is correct providing that such a maneuver is possible. Clearly, this is the usual case for sparse scenarios.

Now, we consider a queueing network comprising several nodes and a number of customers with several classes. In our modeling approach, we map the vehicles onto the customers of the queueing network. And we consider three nodes for each street at each direction, corresponding to front part, end part, and middle part of that street (Fig. 1). Also, we correspond the residence time of a vehicle at each part of the street to the service time of the corresponding customer at the corresponding node. Therefore, in the queueing network, when a customer departs from a node representing the front or middle part of a typical street it is routed definitely to a specific node, i.e., the node representing the middle or end part of the same street, respectively. However, when a customer departs the node representing the end part of a street, i.e., the corresponding vehicle reaches to an intersection, it is routed to three nodes representing the front parts of other three streets at that intersection with some probabilities (Fig. 1).

As we discussed we consider different speed categories at different parts of the streets, representing approximately different behavior of the vehicles along a street. In the proposed mobility model we simply map each speed category onto a customer class. Therefore, we have three customer classes at nodes representing the middle parts of the streets, and two customer classes for the nodes representing the front and end parts of the streets. Each customer class at each node has a random service time with an arbitrary distribution equivalent to the distribution of the corresponding speed category at the corresponding street. Then, the average service time of a customer class at each node is determined with respect to the distribution of its corresponding speed category.

Therefore, we consider each node at the above queueing network, as an $M/G/\infty$ node. So, in the case all parts of the street are in sparse situation, we actually have a queueing network comprised of $M/G/\infty$ nodes with 2 or 3 customer classes, corresponding to middle or front/end parts, respectively. This is a BCMP queueing network with a product-form solution. In this network the following traffic equations are satisfied [10]:

$$\alpha_{ju} = \lambda_{ju} + \sum_{k=1}^{N} \sum_{v=1}^{C(k)} \alpha_{kv} r_{kv,ju}; j = 1, \ldots, N; u = 1, \ldots, C(j) \tag{1}$$

where $\lambda_{ju}$ is the exogenous arrival rate of class-$u$ customers at node $j$, $C(j)$ is the number of classes at node $j$, and $r_{kv,ju}$ denotes the routing probability of a departed class-$v$ customer from node $k$ to node $j$ as a class-$u$ customer. Moreover, $\lambda_{ju}$ equals zero for the internal streets (see Fig. 1). Also, $\lambda_{ju}$ in (1) denotes the arrival rate of class-$u$ customers at node $j$ in the network, representing exogenous arrival rates as well as routing from other nodes. It is obviously equals the departure rate of class-$u$ customers from node $j$ in the case of stability. We also have the following relation for routing probabilities [10]:

$$\sum_{k=1}^{N} \sum_{v=1}^{I} r_{ju,kv} + r_{ju,0} = 1; j = 1, \ldots, N; u = 1, \ldots, I \tag{2}$$

where 0 denotes the world outside the network. Then, we obtain the spatial traffic distribution corresponding to the proposed BCMP queueing network comprised of $M/G/\infty$ nodes, as in the following [10]:

$$\begin{aligned}
\pi(\mathbf{n}) &= \prod_{j=1}^{N} \pi_j(\mathbf{n}_j); \\
\mathbf{n} &= (\mathbf{n}_1, \mathbf{n}_2, \ldots, \mathbf{n}_N), \mathbf{n}_j = (n_{j1}, \ldots, n_{jC(j)}), \\
\pi_j(\mathbf{n}_j) &= e^{-\rho_j} \prod_{u=1}^{C(j)} \frac{\rho_{ju}^{n_{ju}}}{n_{ju}!}; \rho_{ju} = \frac{\alpha_{ju}}{\mu_{ju}}, \alpha_j = \sum_{u=1}^{C(j)} \rho_{ju}.
\end{aligned} \tag{3}$$

Where $\mu_{ju}$ denotes the inverse of average residence time, i.e., service rate, of a class-$u$ customer at node $j$, and $C(j)$ denotes the number of classes at node $j$. In (3), $\pi(\mathbf{n})$ represents the probability of $n_{11}$ class-1 customers at node 1, $n_{12}$ class-2 customers at node 1, etc. By (3) we obtain the probability of existing different number of customers of different speed categories at different parts of the streets, i.e., spatial traffic distribution. Then, by solving the traffic equations and finding the arrival rates of vehicles at each street, we are able to find the probability corresponding to different number of vehicles at that street. We employ this result in the next section for computing the probability of connectivity.

## III. ANALYSIS OF CONNECTIVITY

In this section we compute the probability of connectivity of a street. As it is mentioned before our city is modeled by a square graph and each street is divided to three sections. And also, it turns out from previous discussions that the number of nodes at each section is a Poisson random variable. Let $\rho_1$, $\rho_2$ and $\rho_3$ denote the parameters of the Poisson random variables corresponding to a typical three sections of the street. Since the length of first and last sections are comparable with transmission ranges (see table I), it is just necessary to have at least one node in these sections to be sure of connectivity. We state similar arguments for connectivity of middle section as in [11] and we extend it, in order to find a lower bound for connectivity corresponding to heterogenous transmission ranges. Thus, we have the following relations:

$$
\begin{aligned}
P(con) &= P(con\_sec_1) \times P(con\_sec_2) \times P(con\_sec_3) \\
&= P_{sec_1}(Nodes \neq 0) \times P(con\_sec_2) \times P_{sec_3}(Nodes \neq 0) \\
&\quad (1 - e^{-\rho_1}) \times P(con\_sec_2) \times (1 - e^{-\rho_3}),
\end{aligned}
\tag{4}
$$

where $P(con)$, $P(con\_sec_i)$, $P_{sec_i}(Nodes \neq 0)$ denote the probability of connectivity along the street, probability of connectivity at section $i$ of the street and probability that section $i$ contains at least one vehicle, respectively. Using the results of [11] for one-dimensional connectivity in and considering that the length of the second section of the typical street and transmission ranges are $D$ and $R$, respectively, we conclude that:

$$
\begin{aligned}
P(con\_sec_2) &= \sum_{n=0}^{\infty} P(Nodes = n) \times P(con|_{Nodes=n}) \\
&\sum_{n=0}^{\infty} \frac{e^{-\rho_2} \rho_2^n}{n!} \sum_{i=0}^{m} (-1)^i \binom{n+1}{i} (1 - i\frac{R}{D})^n,
\end{aligned}
\tag{5}
$$

where $m = min\{n+1, \lfloor \frac{D}{R} \rfloor\}$, for each $n$ in the summation.

| Parameter | Value |
|---|---|
| Low speed in the middle part | Uniform $[3, 14]$ $m/s$ |
| Medium speed in the middle part | Uniform $[14, 22]$ $m/s$ |
| Fast speed in the middle part | Uniform $[22, 33]$ $m/s$ |
| Low speed in the front part | Uniform $[0.3, 3]$ $m/s$ |
| Medium speed in the front part | Uniform $[3, 14]$ $m/s$ |
| Low speed in the end part | Uniform $[0.3, 1.5]$ $m/s$ |
| Medium speed in the end part | Uniform $[1.5, 14]$ $m/s$ |
| Length of the middle part | $1600$ $m$ |
| Length of the front part | $200$ $m$ |
| Length of the end part | $200$ $m$ |
| Transmission range | $200$ $m$ |

TABLE I
TYPICAL VALUES USED FOR PARAMETERS OF OUR ANALYSIS

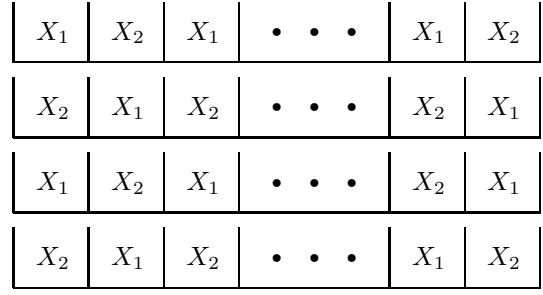

Fig. 2. Four possible states for inserting $r$ nodes of type 2 into q blocks.

Let us assume that we have a street of length $D$ and every vehicle has a transmission range equal to $x_1$ and $x_2$ ($x_1 < x_2$), with the probabilities $p$ and $(1-p)$, respectively. We mention nodes having $x_1$ as a transmission range as nodes of "type 1" and others as nodes of "type 2". Besides, let $\binom{n}{i}$ be zero wherever $i > n$ or $i < 0$ in the following discussions.

*Theorem 3.1:* Suppose we have $N$ nodes distributed independently and uniformly in the street. The probability of connectivity in the street is lower bounded by:

$$
\sum_{r=0}^{N} p^r (1-p)^{N-r} \tilde{Q}(r, N),
\tag{6}
$$

where,

$$
\begin{aligned}
\tilde{Q}(r, N) = \\
\sum_{q=1}^{N} \binom{r-1}{r-q} \binom{N-r+1}{N-r-q+1} p(N+1-r+q, r-q),
\end{aligned}
\tag{7}
$$

for $r > 0$, in addition $\tilde{Q}(0, N) = p(0, N+1)$ and

$$
p(n_1, n_2) = \\
\sum_{k=0}^{N+1} (-1)^k \sum_{l=0}^{k} \binom{n_1}{l} \binom{n_2}{k-l} (1 - lx_1 - (k-l)x_2)^N,
\tag{8}
$$

where the summation is taken over positive values of $1 - lx_1 - (k-l)x_2$.

*Proof:* Suppose that $U_i$ represent the distance of $i^{th}$ node from the origin. Ordering these random variables (RVs) we obtain:

$$
0 \leq U_{(1)} \leq \ldots \leq U_{(N)} \leq D,
\tag{9}
$$

Define:

$$
\begin{aligned}
\Delta_{i-1} &\triangleq U_{(i)} - U_{(i-1)}, U_{(0)} \triangleq 0, U_{(N+1)} \triangleq D, \\
&\forall i = 1, 2, \ldots, N+1.
\end{aligned}
\tag{10}
$$

Again, considering the results for street connectivity [11], for $c_{i_j} \geq 0$, $\{c_{i_j}\}_{j=0}^{r} \subset \{1, 2, \ldots, N\}$ and $\sum_{j=1}^{r} c_{i_j} \leq D$ we know that $(1 - \sum_{j=1}^{r} \frac{c_{i_j}}{D})^N$ expresses the probability $P\{\Delta_{i_1} \geq c_{i_1}, \ldots, \Delta_{i_r} \geq c_{i_r}\}$. Now, using principle of inclusion and exclusion we attain the following relation:

$$
\begin{aligned}
P\{\Delta_{i_1} &\leq c_{i_1}, \ldots, \Delta_{i_r} \leq c_{i_r}\} \\
&= \sum_{j=0}^{r} (-1)^j \sum_{i_1, \ldots, i_j} P\{\Delta_{i_1} \geq c_{i_1}, \ldots, \Delta_{i_j} \geq c_{i_j}\} \\
&= \sum_{j=0}^{r} (-1)^j \sum_{i_1, \ldots, i_j} (1 - \sum_{l=1}^{j} \frac{c_{i_l}}{D})^N.
\end{aligned}
\tag{11}
$$

We take this formula as our template for the corresponding calculation. We shall distinguish nodes with different ranges and their positions in order to find our convenient result. For a moment, we restrict ourselves for a moment to a specific case in which nodes having $x_2$ as transmission range are arranged in a way that there are only $q$ blocks of consequent nodes of type 2. And also assume that only $r$ nodes are of type 2. It is worth mentioning that this case happens with the probability $p^r(1-p)^{N-r}$. By previous assumption, there are only four cases for locating these $q$ blocks and Fig. 2 depicts them. In case 1 and 2, nodes of type 1 are also located in $q$ blocks whereas they are located in $q+1$ and $q-1$ blocks in respect to case 3 and 4.

In Fig. 2, $X_i$ denotes a block of consequent nodes of transmission range $x_i$. Notice that each state has just $q$ blocks including nodes of type 2. In order to compute the number of situations related to case 1 in Fig. 2, we have to put $r$ nodes of type 2 into q blocks and other $N-r$ nodes of type 1 into other $q$ blocks which is given by the equation $\binom{r-1}{r-q}\binom{N-r-1}{N-r-q}$.

To be sure of connectivity, it suffices for the distance of nearby nodes ($\Delta_i s$) to be less than the transmission ranges of those nodes. And also, in order to have connectivity with the entrance and exit part of the street, $\Delta_1$ and $\Delta_{N+1}$ should be less than $x_{(1)}$ and $x_{(N)}$, respectively. But there also exists cases in which street is connected and the distance between some consequent nodes is greater than one of the nearby nodes' transmission range. For example, this can occur when a node of type 1, lying between two connected nodes of type 2, is isolated from them. Thus, as we do not consider these cases, our assertion is a lower bound for connectivity. Next, we choose $c_i s$ in (11) to be $x_1$ or $x_2$ so that our aim will be achieved. Assertion (11) is symmetric and linear, thus it only depends on the number of $c_i s$ which are equal to $x_1$. In case 1, among $\Delta_i s$ ($r+q$) distances are $x_2$ and other $N+1-r-q$ ones are $x_1$. This symmetric property and similar arguments for other cases shed light into the following equation (after replacing $c_i s$ by $x_1$ or $x_2$, (11) will change to (8)):

$$\sum_r p^r(1-p)^{N-r}Q(r,N) \leq P(con), \qquad (12)$$

where,

$Q(r,N) =$
$$\sum_{q=1}^{N}\binom{r-1}{r-q}\binom{N-r-1}{N-r-q-1}p(N+1-r+q, r-q)$$
$$+2\sum_{q=1}^{N}\binom{r-1}{r-q}\binom{N-r-1}{N-r-q}p(N-r+q, r-q+1)$$
$$+\sum_{q=1}^{N}\binom{r-1}{r-q}\binom{N-r-1}{N-r-q+1}p(N-1-r+q, r-q+2),$$
$$\qquad (13)$$

for $r > 0$ and $Q(0,N) = p(0, N+1)$. One can ignore the differences between $p(N+1-r+q, r-q), p(N-r+q, r-q+1)$ and $p(N-1-r+q, r-q+2)$ and approximate (13) by (7). This approximation is on the behalf of obtaining a lower bound. And this completes our proof.

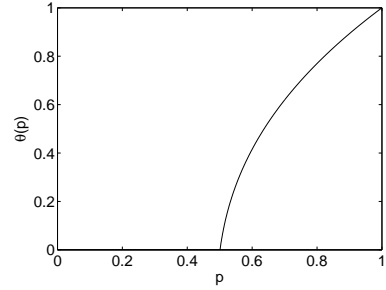

Fig. 3. Probability of bond percolation

It is clear that the inequality reduces to equality for $p = 0$, i.e., one-dimensional connectivity with fixed transmission range. ∎

## IV. A REVIEW ON PERCOLATION THEORY

Percolation theory was originally developed by Hammersley and Broadbent [12] in 1957 as a model for describing fluid spread through porous mediums. It basically deals with the formation of long-range connectivity in random systems. Later, it found many applications in a wide range of phenomena that exhibit phase transitions and critical behavior especially in statistical physics and materials science. More recently, concepts of percolation theory have been applied in the study of the network topology, connectivity and robustness to yield valuable results.

Consider an infinite $2D$ square lattice grid in which each of vertices or edges are independently occupied (open) with probability $p$ and empty (close) with probability $1-p$. The former case is called site percolation and the latter is known as bond percolation. We will restrict our discussion to bond percolation problem, however, similar arguments hold for the site percolation case as well.

For the bond percolation case two open bonds are connected if they share a common vertex or there exists a path of subsequent open bonds between them. Clusters are then formed through connections between nearest neighbor occupied bonds. The network is said to percolate if the size of cluster containing a given vertex, e.g., origin, is infinite. More formally the percolation probability function is defined as follows [13,14]:

$$\theta(p) = Pr_p(|C| = \infty)$$

where $C$ is the cluster containing the origin.

Fig. 3 shows a typical plot of the percolation probability. It is a fundamental result of percolation theory that there exists a critical probability threshold [13] for which

$$\begin{cases} \theta(p) = 0 & \text{for } p < p_c \\ \theta(p) > 0 & \text{for } p > p_c \end{cases}$$

where,

$$0 < p_c = sup\{p : \theta_p = 0\} < 1$$

The value of critical probability depends on the lattice structure, dimensionality of the grid and type of percolation system. Unfortunately the exact values of the percolation thresholds are known for only a few special cases. The following non-trivial theorem for bond percolation threshold in square lattices will be enough for our purpose.

*Theorem 4.1:* The critical probability of bond percolation on the square lattice is 1/2.

Another important result from percolation theory is the zero-one law for the existence of infinite open cluster and its uniqueness.

*Theorem 4.2:* The Probability that an infinite open cluster exists is 0 if $p < p_c$ and 1 if $p > p_c$. The infinite cluster is almost surely unique whenever it exists.

This infinite cluster is also called spanning cluster since it spans the entire lattice. When the value of bond probability is relatively small, the grid mainly consists of isolated islands of nodes with a few pairs of neighboring bonds. As the bond probability increases, these islands grow and some of them merge to form larger clusters. When the occupancy probability is about to exceed the critical threshold, suddenly all clusters are absorbed into one dominating cluster and the unique infinite cluster appears. Above this threshold, other small size clusters gradually join the spanning cluster. So clearly $\theta(p)$ is exactly zero in the subcritical phase, since there is no spanning cluster in this phase.

For the finite size lattices, however, instead of a single point there exists a transition region where the percolation probability raises sharply with the increase in the value of edge probability. Here the transition region becomes narrower with the growth of lattice size. We will discuss this matter later in numerical results.

## V. Computational Algorithm

In this section we will briefly introduce an efficient Monte-Carlo approach estimating desired parameters in percolation systems. It is clear that testing all permutations of a bond percolation problem in a $N \times N$ grid is computationally intractable, since there exists $2N(N-1)$ bonds and that the solution requires $O(2^{2N(N-1)})$ of time even for a single value of edge probability. Therefore, using a Monte Carlo approach seems to be inevitable.

Newman and Ziff [15] proposed a fast algorithm for the estimation of percolation probabilities and lattice quantities directly for any arbitrary value of edge probability. The idea is that in order to estimate any parameter, e.g. average or giant cluster size, in a finite size percolation it suffices to only estimate that parameter for all possible numbers of total edges in the corresponding lattice and then use the law of total probability:

$$Q(p) = \sum_m P(M=m)Q_m = \sum_m \binom{M}{m} p^m (1-p)^{(M-m)} Q_m$$
(14)

where M denotes the total number of edges in the lattice and $Q_m$ represents the expected value of the desired parameter in a lattice with $m$ total occupied bonds.

Now the problem reduces to the computation of $Q_m$ coefficients. Consider for example "average cluster size" and "giant cluster size" as two parameters of interest. So we need to find contiguous clusters of open bonds. One simple suggestion for the estimation of these parameters is to independently generate a number of configurations of the percolation system over all possible states with total $M$ edges. The clusters can then be obtained using a simple depth-first search (DFS) or breadth-first Search (BFS). However one can easily check that this approach is not computationally effective and is intractable for large values of lattice size [15].

We can do much better by noting that any state with $n+1$ bonds is generated by adding an extra randomly chosen bond to a sample state with $n$ bonds. Thus we can start with an empty lattice and add one bond to it at each step. The advantage is that we can keep track of the clusters and update them at each step within much less operations since each bond can merge two clusters at most.

By repeating this procedure for a number of iterations and averaging over the sample lattice quantities we expect to have a good estimate of these values. This is because the precision of a statistical measurement depends on the number of randomly samples taken and not the size of the sample space.

To keep track of the clusters and update them at each step we need to find an efficient data structure to perform two operations efficiently. These operations are Find(x) that returns the cluster that node x belongs and Union (x, y) to merge two clusters that contain x, y nodes whenever we add a bond between them. This problem has been extensively studied in computer science and there exists a family of efficient algorithms for it, generally known as Union-Find Algorithms. We can summarize this algorithm for bond percolation problem in an $N \times N$ square grid in the following steps:

- Initialization: Consider each of $N^2$ nodes in an empty lattice as single clusters. Each node will have a pointer to its parents, initially all the pointers will be null.
- Find(x): We start from x, and recursively traverse up the tree following the parent pointers till we get to the root which will be a node with null parent pointer. Return the root as the cluster which contains x. Each time this procedure is called we set all the parent pointers of all the nodes on the path to the root directly to the root of the tree (Path compression step).
- Union(x, y): First we perform two find operations to obtain the corresponding cluster roots of each node. If both nodes belong to a single cluster we do nothing, otherwise we merge them by setting the parent pointer of smaller cluster root to the parent pointer of the larger cluster's root. Thus we should anly keep track of the cluster sizes for each of the roots.

Analysis of this algorithm [15] shows that performing a sequence of $m$ operations over $n$ elements requires an overall

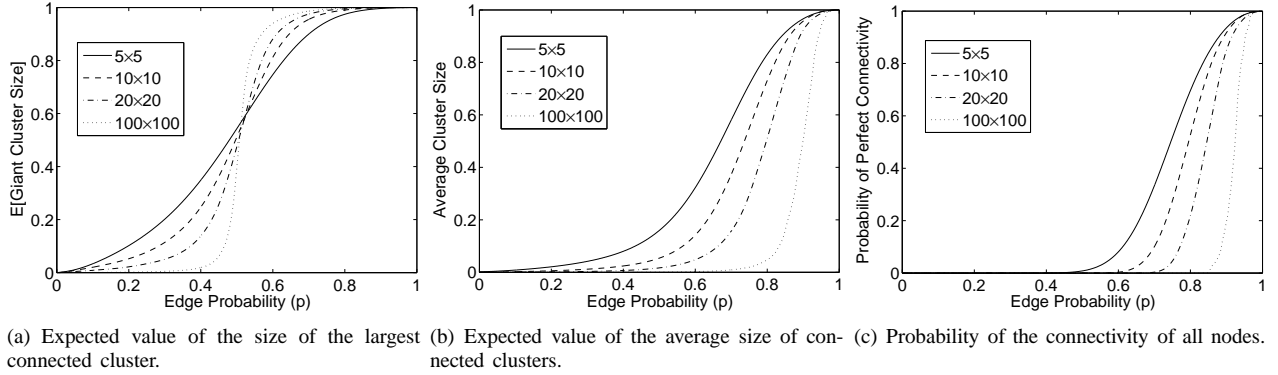

(a) Expected value of the size of the largest connected cluster.

(b) Expected value of the average size of connected clusters.

(c) Probability of the connectivity of all nodes.

Fig. 4. Connectivity properties of finite bond percolation as a function of edge probability, for different square sizes.

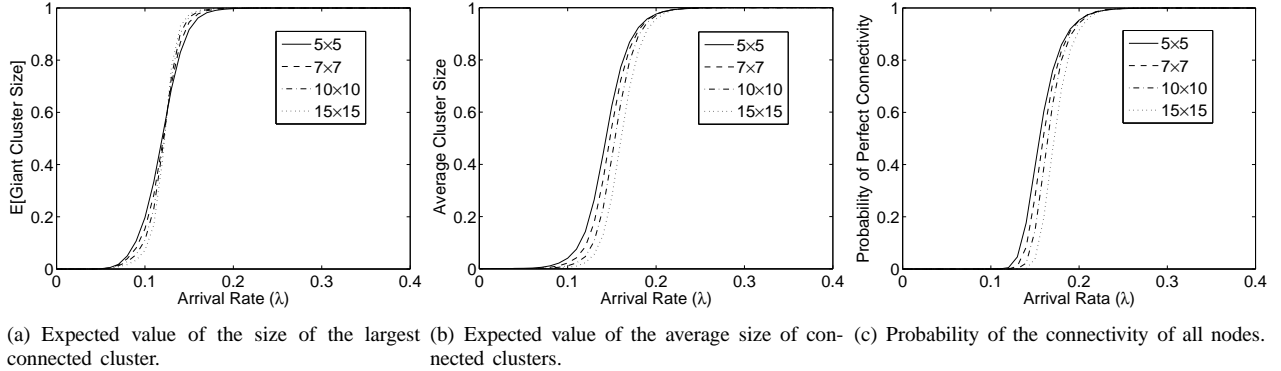

(a) Expected value of the size of the largest connected cluster.

(b) Expected value of the average size of connected clusters.

(c) Probability of the connectivity of all nodes.

Fig. 5. Connectivity properties of VANETs as a function of entrance rate of vehicles, for different city sizes.

running time of $O((n+m)\log^*(n))$ which grows almost linear with the size of grid nodes in practice. It should be noted that this algorithm can be easily modified to find percolation thresholds and different properties for other lattice shapes and percolation systems as well.

## VI. NUMERICAL RESULTS AND SIMULATIONS

In this section, we will investigate our approach to analyse connectivity features of the VANETs in a number of scenarios. We will consider three measures for the connectivity: "Giant Cluster Size" which is the fraction of the intersections that belong to the largest connected cluster, "Average size of connected clusters" or simply "Average Cluster Size" and finally "Probability of Perfect Connectivity", i.e., the probability that all intersections are connected to each other. In all our scenarios, we will use typical values shown in Table I, for different parameters of the presented mobility model (section II).

### Scenario 1. Connectivity as a function of Edge Probability

First, we examine connectivity properties of the finite bond percolation problem as a function of edge probability (p). In our model of the city, this will correspond to the connectivity of the intersections of the city as a function of the street connectivity. These parameters are illustrated in Fig. 4 for different lattice sizes. The emergence of giant cluster size at about the edge probability of $1/2$ completely agrees with theorem 4.1 (Fig.4(a)). As expected, this transition is sharper for larger values of lattice size. In Fig. 4(b), 4(c), we observe that the transition for average cluster size and probability of perfect connectivity occurs in a larger value of edge probability depending on the lattice size. This is due to the fact that, even after the appearance of a unique giant cluster (theorem 4.2), there are still many isolated islands of nodes in the lattice. It is not also surprising to see the increase of the transition probability for larger values of lattice size.

### Scenario 2. Connectivity as a function of Entrance Rate

We demonstrate the results of our analysis of connectivity properties as a function of the entrance rate of the vehicles. We use our queueing network model to find the probability of street connectivity and then apply the fast algorithm proposed in section V to find the desired parameters. Fig. 6 shows the probability of a street connectivity as a function of entrance rate of the vehicles (3), (4). The connectivity properties, previously mentioned, are demonstrated in Fig. 5.

It is also necessary to validate the analytical results of our method with the aid of a computer simulation. With the same parameter setup, the simulation results are compared with analysis results in Fig. 7. The congruence of these diagrams is approved our method of connectivity analytical for 2D vehicular networks.

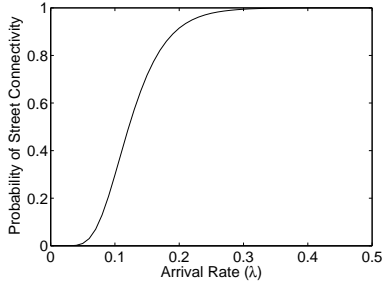

Fig. 6. the probability of connectivity of an arbitrary street in a $7 \times 7$ symmetric city

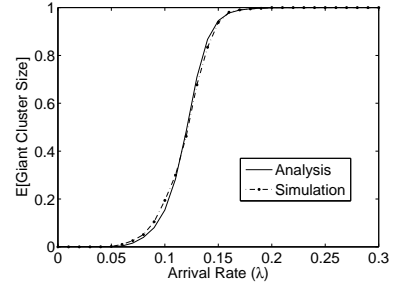

Fig. 7. Simulation and analysis results for Giant Cluster Size versus Entrance Rate($\lambda$) in a $7 \times 7$ symmetric city

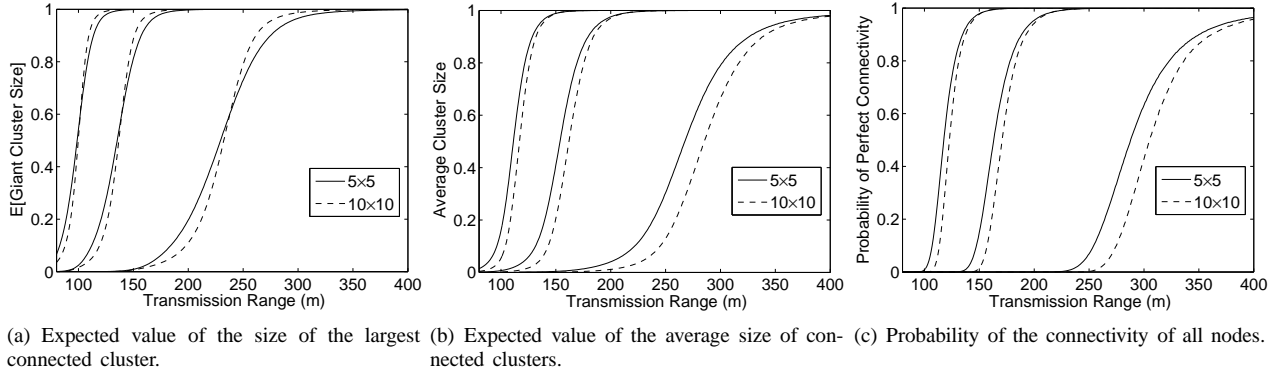

(a) Expected value of the size of the largest connected cluster.

(b) Expected value of the average size of connected clusters.

(c) Probability of the connectivity of all nodes.

Fig. 8. Connectivity properties of VANETs as a function of transmission range of Vehicles, for different city sizes.

### Scenario 3. Effect of Transmission Range

As an application of our method, we now consider the effect of the transmission range of the vehicles on the connectivity properties of the network. The problem of finding optimal range for radio transmission is one of the main concerns of the wireless design engineers, since choosing an inappropriate transmission range, whether low or high, is destructive for the network. The disadvantage of low-power transmission is that we obviously lose the network connectivity. On the other hand, a very high power of transmission is not only a waste of energy, but also a cause of interference among nodes in the network. Fig. 8 displays the connectivity properties of the VANETs with respect to the transmission ranges of the vehicles for several values of arrival ranges, i.e., $\lambda = 0.1, 0.2, 0.3$ respectively. Clearly, given an estimation of entrance rate of a city, we can find the optimal value for the transmission range of vehicles.

### Scenario 4. VANET Connectivity in an asymmetric city

In our mobility model, we considered a symmetric model for the routing of vehicles between distinct intersections. However, in a real situation this assumption is not valid because different parts of the city have certainly different loads of traffic. We can simply generalize our mobility model to solve this problem by assigning a traffic weight factor to each intersection of the city. The vehicles will then be routed between intersections according to this factor, i.e., the probability that a vehicle chooses one intersection as its next destination will be proportional to the traffic weight of the corresponding intersection.

In this case, the street connectivity probabilities of the city are not equal any more. However, note that we can find a lower and upper bound for connectivity characteristics by assuming that the probability of all streets equal to the least and largest probabilities among all streets of the city, respectively. As an example, we consider a city in which all traffic weights are distributed uniformly in [1,2]. Fig. 9 illustrates these bounds for a sample city with such a traffic distribution model mentioned above. We are much more interested in the lower bounds, since upper entrance rates will guarantee the network connectivity. The upper bounds gives us a rough estimation of the precision of our lower bound.

### Scenario 5. Multiple Transmission Ranges

Finally, we take our theoretical results of section III into account for the problem of street connectivity in a network of vehicles with two transmission ranges. Here, each node is given transmission range $200m$ or $400m$ with probability $p$ or $(1 - p)$ respectively. Obviously, every street connected by nodes having transmission range less than or equal to $2R$ will be connected if we set all transmission ranges to $2R$. Thus, the graph related to $p = 0$ can be thought of as a upper bound for other graphs. Fig. 10 shows lower and upper bounds provided by Theorem 3.1.

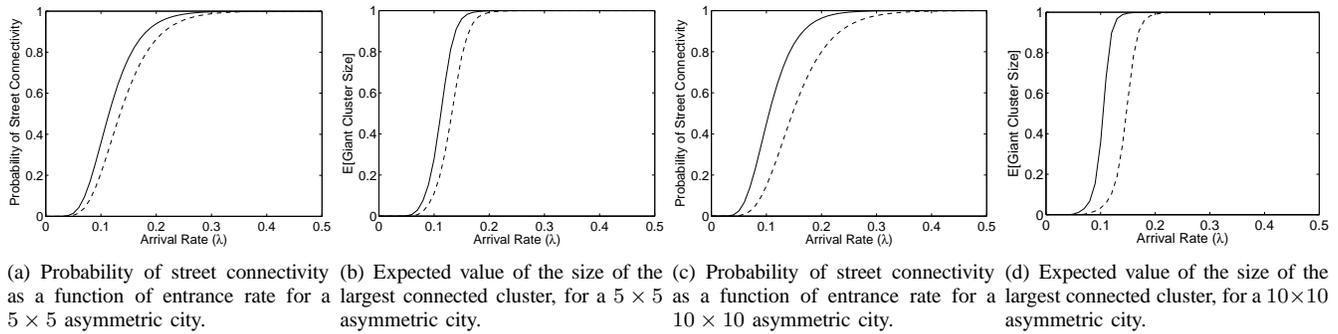

(a) Probability of street connectivity as a function of entrance rate for a $5 \times 5$ asymmetric city.

(b) Expected value of the size of the largest connected cluster, for a $5 \times 5$ asymmetric city.

(c) Probability of street connectivity as a function of entrance rate for a $10 \times 10$ asymmetric city.

(d) Expected value of the size of the largest connected cluster, for a $10 \times 10$ asymmetric city.

Fig. 9. Bounds on connectivity properties obtained from minimum and maximum probabilities of street connectivity in a random asymmetric city.

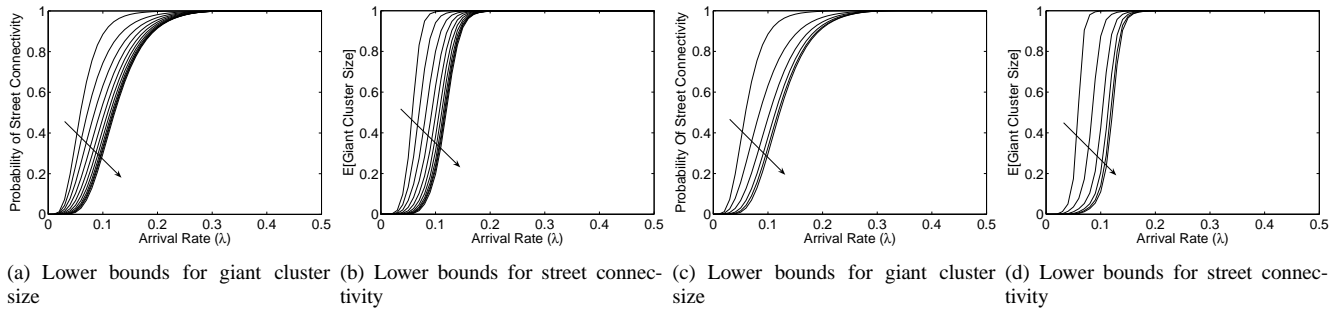

(a) Lower bounds for giant cluster size

(b) Lower bounds for street connectivity

(c) Lower bounds for giant cluster size

(d) Lower bounds for street connectivity

Fig. 10. Lower bounds on connectivity properties (section III) for different values of p, the probability that a node takes $x_1 = R$ as its transmission range. ($x_2 = 2R$, $R = 200m$).

## VII. Conclusions

In this paper, we considered two dimensional connectivity of VANETs and proposed a new method for evaluating connectivity measures such as average size of the connected intersections and size of the giant cluster of them, as a function of the entrance rates of vehicles. Our approach was based on modeling the mobility of the vehicles with a BCMP open queueing network to find spatial distribution of the vehicles and exact probability of connectivity for each street. We then exploit results of percolation theory and a fast algorithm for computing and interpreting connectivity properties of the large scale network. We also derived lower bounds on the connectivity probability of each street for the heterogeneous case in which vehicles could have two different transmission ranges. Finally we illustrated the efficiency of our approach in a number of scenarios, each yielding to some valuable results. These included finding bounds on connectivity features in a city with asymmetric traffic load and determining optimal value for the transmission range of vehicles for a given entrance rate of a city.

## References

[1] Fan Li, Yu Wang, "Routing in Vehicular Ad Hoc Networks: A Survey," *IEEE Vehicular Technology Magazine*, Volume : 2, Issue: 2, Pages: 12-22, June 2007.

[2] J. Zhao and G. Cao, "VADD: Vehicle-Assisted Data Delivery in Vehicular Ad Hoc Networks," *IEEE Proc. INFOCOM06*, 2006.

[3] Madhav Desai, D. Manjunath, "On the Connectivity in Finite Ad Hoc Networks," *IEEE Communications Letters*, vol 10, no 6, pp 437-490, Oct 2002.

[4] G. H. Mohimani, F. Ashtiani, A. Javanmard, M. Hamdi, "Mobility Modeling, Spatial Traffic Distribution, and Probability of Connectivity for Sparse and Dense Vehicular Ad Hoc Networks," *IEEE Transactions on Vehicular Technology*, Accepted for publication.

[5] C. H. Foh, G. Liu, B. S. Lee, B.-C. Seet, K.-J. Wong, C. P. Fu, "Network Connectivity of One-Dimensional MANETs with Random Waypoint Movement," *IEEE Communications Letters*, vol. 9, no. 1, pp. 31-33, January 2005.

[6] C. Bettstetter, "On the connectivity of wireless multihop networks with homogeneous and inhomogeneous range assignment,", *Proc. IEEE*, 2002, pp. 1706-1710.

[7] Tae-Suk Kim, Seong-Lyun Kim, "Random power control in wireless ad hoc networks," *IEEE Communications Letters*, vol. 9, no. 12, pp. 1046-1048, Dec. 2005.

[8] O. Dousse, F. Baccelli and P. Thiran, "Impact of interferences on connectivity of ad hoc networks," *IEEE/ACM Transactions on Networking*, Vol. 13, Nr. 2, pp. 425-436, 2005.

[9] M. K. Mehmet Ali and M. Khabazian, "A Performance Modeling of Connectivity in Ad Hoc Networks," *Proc. WCNC 2006*, paper NET17-1.

[10] X. Chao, M. Miyazawa, and M. Pinedo, *Queueing Networks: Customers, Signals and Product Form Solutions*, John Wiley, 1999.

[11] A. Ghasemi, S. Nader-Esfahani, "Exact probability of connectivity in one-dimensional ad hoc wireless networks," *IEEE Communications Letters*, vol. 10, pp. 251-253, Apr 2006.

[12] S. R. Broadbent, J. M. Hammersley, "Percolation processes I: Crystals and mazes." *Proc. Camb. Phil. Soc.* 53, 629-641, (1957).

[13] G. Grimmett, *Percolation*, 2nd ed., Springer-Verlag, Berlin, 1999.

[14] B. Bollobas, O. Riordan, Percolation. Cambridge University Press, Cambridge, 2006.

[15] M. E. J. Newman, R. M. Ziff, "A fast Monte Carlo algorithm for site or bond percolation," *Phys. Rev. E* 64, 016706 (2001).